\documentclass[12pt, a4paper]{article}%%%%where rsproca is the template name

\usepackage{amssymb,amsmath,mathrsfs,subfigure,epsfig,color}
\usepackage{amsmath}
\usepackage[latin1]{inputenc}
\usepackage{geometry} % to change the page dimensions

\def\d{\mathrm{d}}
\def\im{\mathrm{i}}

\def \div{\mbox{div\hskip 1pt}}

\def \tr{\mbox{tr\hskip 1pt}}
\def \grad{\mbox{grad\hskip 1pt}}

\begin{document}

%%%% Article title to be placed here
\title{Straightening: Existence, uniqueness and  stability}

\author{%%%% Author details
M. Destrade$^{1,2}$, R.W. Ogden$^{3}$, I. Sgura$^{4}$ and L. Vergori$^{1,3}$
\\[24pt]
$^1$School of Mathematics, Statistics and Applied Mathematics,\\
NUI Galway, \\ University Road, Galway, Ireland.\\[12pt]
$^2$School of Mechanical and Materials Engineering, \\
University College Dublin,\\ Belfield, Dublin 4, Ireland.\\[12pt]
$^3$School of Mathematics and Statistics, \\
University of Glasgow, \\
University Gardens,
Glasgow G12 8QW,
Scotland, UK.\\[12pt]
$^4$Dipartimento di Matematica e Fisica ``Ennio De Giorgi",\\ Universit\`a del Salento, Lecce, Italy.}

\date{}
\maketitle

%%%% Keyword entries to be placed here %%%%
%%%% Abstract text to be placed here %%%%%%%%%%%%

\vspace{24pt}

\begin{abstract}

One of the least studied universal deformations of incompressible nonlinear elasticity, namely the straightening of a sector of a circular cylinder into a rectangular block, is revisited here and, in particular, issues of existence and stability are addressed. Particular attention is paid to the system of forces required to sustain the large static deformation, including by the application of end couples. 
The influence of geometric parameters and constitutive models on the appearance of wrinkles on the compressed face of the block is also studied. 
Different numerical methods for solving the incremental stability problem are compared and it is found that the impedance matrix method, based on the resolution of a matrix Riccati differential equation, is the {\color{black}more} precise.
\end{abstract}

\vspace{24pt}

\emph{keywords: nonlinear elasticity, straightening, instability, stiffening}

\newpage

\section{Introduction}

%%%%%%%%%%%%%%%%%%

The rubber of a car tyre in contact with the road is slightly flattened, or \emph{straightened}, with respect to its natural unloaded configuration. 
In other words, a portion of the rubber undergoes a deformation which can be quite accurately captured by Ericksen's solution \cite{Eric54} for the elastic straightening of a circular cylindrical sector into a rectangular block. 
Other examples of application for this deformation include the local behaviour of rubber-covered rollers in service or of extended body joints such as knees and elbows.
Ericksen's exact solution is one of only a handful of universal deformations in incompressible isotropic nonlinear elasticity \cite{TrusNoll}, but it has so far received scant attention in the literature, beyond the works of Hill \cite{Hill73}, Aron and co-workers \cite{Aron98, Aron00, Aron05}, and our recent contribution \cite{DOSV13}. In this paper we complete the picture with some additional results for the (plane strain) large straightening deformation of an incompressible isotropic sector and its stability with respect to incremental deformations.

In \S \ref{sec-2}, we describe the considered deformation, the parameters it involves, and the different boundary conditions under which it can be achieved, illustrated by use of the neo-Hookean strain-energy function. In \S \ref{sec-3}, following a brief discussion of strong ellipticity  of an incompressible isotropic strain-energy function, it is shown that if the sector is straightened \emph{either} by the application of end couples alone \emph{or}, in the absence of end couples, by lateral normal forces alone, then, under the inequalities associated with the strong ellipticity condition, existence of the straightening deformation is guaranteed irrespective of the particular form of strain-energy function.  For a thin sector, asymptotic formulas in terms of the thickness to (outer) radius ratio, denoted $\varepsilon$, are then provided for these two cases to give explicit results for certain parameters of the problem.  In particular, it is found that to third order in $\varepsilon$ the results are independent of the choice of energy function.

In \S \ref{sec-4} we derive the equations of incremental equilibrium in the Stroh form with a view to solving them numerically in order to investigate the possible appearance of wrinkles (i.e. small amplitude undulations or instabilities) at a critical threshold of deformation.  Then, in \S \ref{sec-5}, the equations are effectively solved numerically for the corresponding, numerically stiff, two-point boundary value problem. First we use the Compound Matrix method, which must be slightly modified from its usual form to circumvent a singularity problem. Impedance Matrix techniques are then applied, and this approach proves to be {\color{black}more} precise for the problem at hand.
The results are illustrated for homogeneous sectors made of neo-Hookean and of Gent materials, and the effects of geometrical and constitutive parameters on stability are highlighted. 

%%%%%%%%%%%%%%%%%%%%%%%

\section{Basic equations\label{sec-2}}

%%%%%%%%%%%%%%%%%

\begin{figure}
	\centering
		\includegraphics[  width=0.9\textwidth]{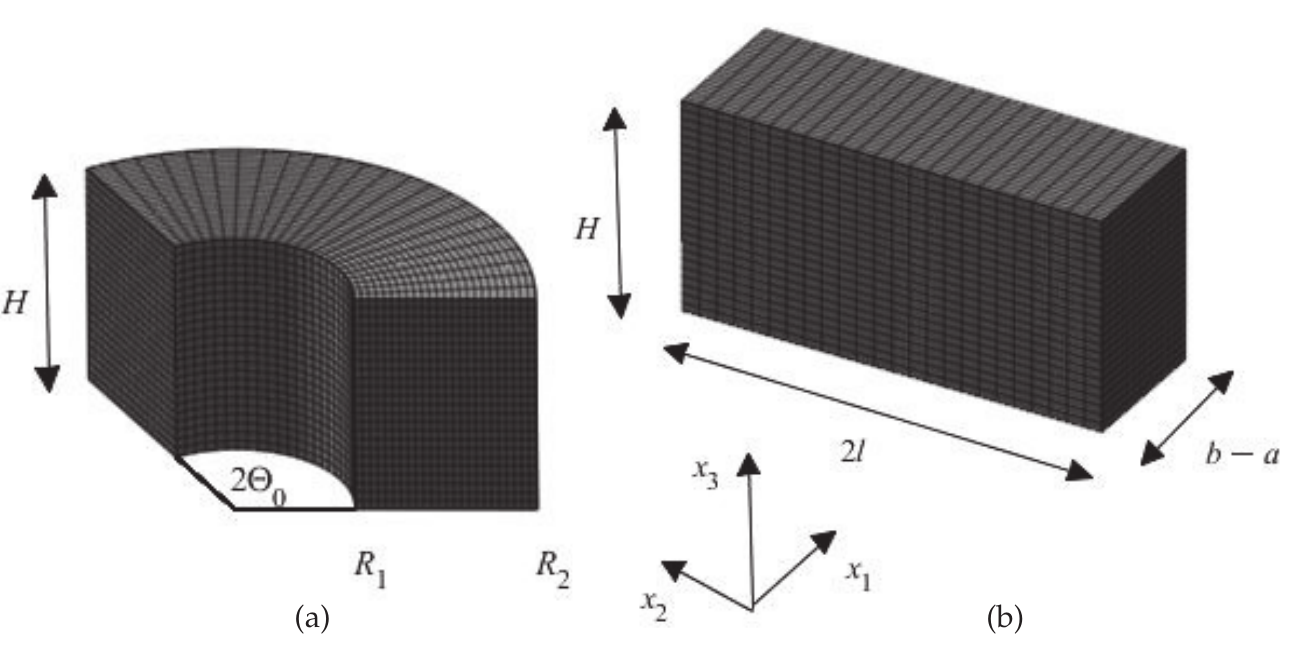}	\label{fig:figure1}
		\caption{(a) A circular cylindrical sector with internal and external radii $R_1$ and $R_2$, respectively, and sector angle $2\Theta_0$ straightened (b) under plane strain conditions into a rectangular block of thickness $b-a$ and length $2l$.\label{fig1}}
\end{figure}

Consider the circular cylindrical sector of incompressible isotropic elastic material shown in figure \ref{fig1}(a) in terms of cylindrical polar coordinates $(R, \Theta,Z)$, with geometry defined by the reference region
\begin{equation}
0<R_1\leq R\leq R_2,\quad -\Theta_0\leq \Theta\leq \Theta_0,\quad 0\leq Z\leq H,
\end{equation}
where $0<2\Theta_0<2\pi$ is the angle spanned by the sector.  
The sector can be deformed into a rectangular block, as shown in figure \ref{fig1}(b) with respect to Cartesian coordinates $(x_1,x_2,x_3)$, by the deformation \cite{TrusNoll}
\begin{equation}
x_1=\tfrac{1}{2}AR^2,\quad x_2=\frac{\Theta}{A},\quad x_3=Z,
\end{equation}
where $A=\Theta_0/l$ and $2l$ is the length of the block in the $x_2$-direction. 
Here we are restricting the study to a plane strain deformation, although a uniform stretch could easily be included in the $x_3$-direction \cite{Eric54, TrusNoll}.
The deformed straightened sector occupies the region described by
\begin{equation}
a\leq x_1\leq b,\quad -l\leq x_2\leq l, \quad 0\leq x_3\leq H,
\end{equation}
where $a$ and $b$ are defined as
\begin{equation}
a=\tfrac{1}{2}AR_1^2, \quad b=\tfrac{1}{2}AR_2^2.\label{abR1R2}
\end{equation}

The corresponding deformation gradient  $\mathbf{F}$ has the form
\begin{equation}
\mathbf{F}=AR\mathbf{e}_1\otimes\mathbf{E}_R+\frac{1}{AR}\mathbf{e}_2\otimes\mathbf{E}_\Theta+\mathbf{e}_3\otimes\mathbf{E}_Z,
\end{equation}
where $\mathbf{E}_R,\mathbf{E}_\Theta,\mathbf{E}_Z$ and $\mathbf{e}_1,\mathbf{e}_2,\mathbf{e}_3$ are the cylindrical polar and Cartesian unit basis vectors in the reference and deformed configurations, respectively.
It follows that the Eulerian principal directions of the deformation (defined as the directions of the eigenvectors of the left Cauchy--Green deformation tensor $\mathbf{B}=\mathbf{F}\mathbf{F}^\mathrm{T}$) are the Cartesian basis vectors and that the principal stretches are 
\begin{equation}
\lambda_1=AR,\quad \lambda_2=\frac{1}{AR},\quad \lambda_3=1.\label{lambda1lambda2}
\end{equation} 

We consider an incompressible isotropic hyperelastic material with strain energy $W=W(\lambda_1,\lambda_2,\lambda_3)$ per unit volume, so that the Cauchy stress tensor can be written as
\begin{equation}
\boldsymbol{\sigma}=\sigma_1\mathbf{e}_1\otimes\mathbf{e}_1+\sigma_2\mathbf{e}_2\otimes\mathbf{e}_2+\sigma_3\mathbf{e}_3\otimes\mathbf{e}_3,
\end{equation}
where $\sigma_1,\sigma_2,\sigma_3$ are the principal Cauchy stresses given by \cite{Ogde97}
\begin{equation}
\sigma_1=\lambda_1\frac{\partial W}{\partial \lambda_1}-p,\quad \sigma_2=\lambda_2\frac{\partial W}{\partial \lambda_2}-p,\quad \sigma_3=\frac{\partial W}{\partial \lambda_3}-p,\label{sigmas}
\end{equation}
$p$ being a Lagrange multiplier associated with the incompressibility constraint $\lambda_1\lambda_2\lambda_3=1$, which is automatically satisfied by \eqref{lambda1lambda2}.

Henceforth, it is convenient to use the notation $\lambda_2=\lambda$, $\lambda_1=\lambda^{-1}$, and to introduce the function $\hat{W}$  of a single deformation variable defined by 
\begin{equation}
\hat{W}(\lambda)=W(\lambda^{-1},\lambda,1),
\end{equation}
from which, on use of  \eqref{sigmas}, we obtain
\begin{equation}\label{sigmas diff}
\sigma_2-\sigma_1=\lambda\hat{W}'(\lambda).
\end{equation}

Since the deformation depends only on the single variable $R$ (or $x_1$), the second and third components of the equilibrium equation $\div\boldsymbol{\sigma}=\mathbf{0}$ in the absence of body forces show that $p$ is independent of $x_2$ and $x_3$, and the first component yields simply $\mathrm{d}\sigma_1/\mathrm{d} x_1=0$; hence $\sigma_1$ is a constant.  Then, by taking the boundary $R=R_1$, for example, to be traction free it follows that $\sigma_1\equiv 0$, and hence
\begin{equation}\label{sigma2}
\sigma_2=\lambda\hat{W}'(\lambda).
\end{equation}
If required, the value of $\sigma_3$ needed to maintain the plane strain condition may be obtained in terms of $\lambda$ from \eqref{sigmas}$_3$ with $p=\lambda_1\partial W/\partial\lambda_1$.

Next, we compute the resultant normal force $N$ and moment $M$ (about the origin of the Cartesian coordinate system) on the end  face $x_2=l$ of the block.  They are given by
\begin{equation}
N = H\int_a^b\sigma_2\mathrm{d}x_1,\quad M = -H\int_a^b\sigma_2x_1\mathrm{d}x_1.\label{MN-formulas}
\end{equation}
Note that because $\sigma_2$ is independent of $x_2$, $N$ and $M$ are in fact the same on any section of the block normal to the $x_2$-direction.
By a change of variables we arrive at 
\begin{equation}\label{NM}
N = HR_2\lambda_b\int_{\lambda_b}^{\lambda_a}\frac{\hat{W}'(\lambda)}{\lambda^2}\mathrm{d}\lambda,\quad 
M=-\frac{HR^2\lambda_b^2}{2}\int_{\lambda_b}^{\lambda_a}\frac{\hat{W}'(\lambda)}{\lambda^4}\mathrm{d}\lambda,
\end{equation}
where 
\begin{equation} \label{lambda_ab}
\lambda_a=\frac{1}{AR_1}, \quad \lambda_b=\frac{1}{AR_2}  =\frac{R_1}{R_2}\lambda_a
\end{equation}
are the values of the stretch $\lambda$ on the faces $x_1=a$ and $x_1=b$, respectively, of the straightened block. 
Except for large values of $|N|$ it is expected that the circumferential elements on the ``inner'' face of the straightened block are extended and those on the ``outer'' face are contracted, i.e. $\lambda_a>1$ and $\lambda_b<1$, in which case, by \eqref{lambda_ab} and the definition of $A$, it would follow that $l$ belongs to the interval 
\begin{equation}
R_1\Theta_0 < l < R_2\Theta_0. \label{Ainterval}
\end{equation}
We do not insist that this ordering holds in general, but it turns out that it gives a necessary and sufficient condition for the existence of a {\color{black}plane where $\lambda=1$ (with equation $x_1=l/2\Theta_0$) in the straightened block, which we refer to as the \emph{neutral plane}, or \emph{neutral axis} in the $(x_1,x_2)$ plane}. This is the case when either $M=0$ or $N=0$, the two examples we analyse in \S \ref{sec-3}, if we impose the physically reasonable requirement that the stress $\sigma_2$ be positive (negative) when $\lambda>1\,(<1)$, i.e.
\begin{equation}
\hat{W}'(\lambda)\gtreqqless 0\quad\mbox{according as}\quad \lambda \gtreqqless 1.\label{lambdaineqs}
\end{equation}
These inequalities certainly hold when the strain-energy function $W$ satisfies the strong ellipticity condition. 
Indeed,  by \eqref{sigmas diff} we have
\begin{equation}
\frac{\lambda^2\hat{W}'(\lambda)}{\lambda^2-1}=\frac{\sigma_2-\sigma_1}{\lambda_2-\lambda_1}>0 \quad \textrm{ for } \lambda\neq1,
\end{equation}
because of the Baker--Ericksen inequalities, which are a consequence of the strong ellipticity condition \cite{TrusNoll}. 
Then clearly, \eqref{lambdaineqs} readily follows.

As an example of the large straightening deformation,  we consider the neo-Hookean material, for which
\begin{equation}
W=\tfrac{1}{2} \mu (\lambda_1^2+\lambda_2^2+\lambda_3^2-3),\label{neoH}
\end{equation}
where $\mu >0$ is the ground state shear modulus. For the plane strain problem this reduces to
\begin{equation}
\hat{W}(\lambda)=\tfrac{1}{2}\mu(\lambda^2+\lambda^{-2}-2).
\end{equation}
We then calculate
\begin{equation}
N=\mu HR_2\lambda_b\left[\ln(R_2/R_1)-\frac{1}{4\lambda_b^{4}}\left(1-\frac{R_1^4}{R_2^4}\right)\right]
\end{equation}
and 
\begin{equation}
M=-\frac{1}{4}\mu HR^2_2\left[1-\frac{R_1^2}{R_2^2}-\frac{1}{3\lambda_b^4}\left(1-\frac{R_1^6}{R_2^6}\right)\right].
\end{equation}
Through this explicit example, which as far as we are aware is not available in the literature, it can be seen that in order to describe the straightening deformation, for any given $\Theta_0$, \emph{either} the loads can be prescribed, i.e. $N$ (or $M$) can be prescribed and then the corresponding $A$ and $M$ (or $N$) can be computed from equation \eqref{NM} with \eqref{lambda_ab}, \emph{or} the deformed geometry can be described, i.e. the length $2l$ can be prescribed, hence fixing $A$, and then $N$ and $M$ deduced from equation \eqref{NM} with \eqref{lambda_ab}. In this paper, following considerations of Hill \cite{Hill73} in respect of a spherical cap, we deal with three case studies that are important physically:
\begin{itemize}
\item[({i})] the sector is straightened by end couples alone ($N=0$);
\item[({ii})] the sector is straightened by vice-clamps ($M=0$);
\item[({iii})] with $NM \ne 0$ in general, the final length $2l$ of the straightened sector is determined at the onset of instability.
\end{itemize}

For instance, if the deformation for the neo-Hookean material is achieved by the application of moments alone, as in Case ({i}), then $N=0$ and $A$ is determined  by
\begin{equation}
A^4=\frac{4\ln(R_2/R_1)}{R_2^4-R_1^4},
\end{equation}
in which case $M$ depends on the geometry only through $R_1$ and $R_2$.  

%%%%%%%%%%%%%%%%%%%%%%%%%%%%%%

\section{Examples of straightening\label{sec-3}}

%%%%%%%%%%%%%%%%%%%%%%%%%%%%%%%

This section is concerned with deformations that are achieved by the application of two special systems of forces, that corresponding to zero resultant normal force, $N=0$, and that corresponding to zero resultant moment, $M=0$, i.e. Cases ({i}) and ({ii}) above, respectively.  {\color{black}With reference to \eqref{MN-formulas}, we see that the difference between these two cases arises from the different distributions of the stress $\sigma_2$ with respect to $x_1$ in $[a,b]$.}

We focus on constitutive models that satisfy the strong ellipticity condition, which, for plane strain, consists of the inequalities \cite{Ogde97}
\begin{equation}
\frac{\hat{W}'(\lambda)}{\lambda^2-1}> 0,\quad \lambda^2\hat{W}''(\lambda)+\frac{2\lambda\hat{W}'(\lambda)}{\lambda^2+1}>0.
\end{equation}
These two inequalities are satisfied by many standard strain-energy functions, including the {\color{black}neo-Hookean model:
\begin{equation} \label{MR}
W_\text{nH}=\frac{\mu}{2}(I_1-3) ,
\end{equation}
where $I_1=\tr\mathbf{B}$ and $I_2=\tr(\mathbf{B}^{-1})$ are principal invariants of $\mathbf{B}$ and $\mu$ is a  constant;
the Varga model \cite{Varga}:
\begin{equation} \label{V}
W_\text{V}=2\mu(i_1-3),
\end{equation}
where} $i_1=\tr(\mathbf{B}^{1/2})$; the Fung--Demiray model \cite{Demiray}: 
\begin{equation}\label{F}
W_\text{FD}=\frac{\mu}{2c}\{\exp[c(I_1-3)]-1\}, \quad  c>0,
\end{equation}
where $c$ is a  constant; and the Gent model \cite{Gent96}:
\begin{equation}\label{G}
W_\text{G}=-\frac{\mu J_m}{2}\ln\left(1-\frac{I_1-3}{J_m}\right), \quad J_m>0,
\end{equation}
where $J_m$ is a  constant and the range of deformation is limited by the condition that $I_1-3<J_m$.  In each case $\mu\,(>0)$ is the shear modulus of the material in the undeformed configuration, given by
$\mu= \tfrac{1}{4}\hat W''(1)$.

%%%%%%%%%%%%%%%%%%%%%%%%%%%%%%%

\subsection{Straightening by end couples}\label{couples}

%%%%%%%%%%%%%%%%%%%%%%%%%%%%%%%

The system of loads consists of \emph{end couples} alone when $N=0$, which is the case we consider here.  If $N=0$ then $\sigma_2$ must take both positive and negative signs in the interval $[a,b]$, and, in particular, there must be a value of $x_2$ where $\sigma_2=0$, and hence, by \eqref{lambdaineqs}, $\lambda=1$, so that 
\begin{equation}\label{l1geql2}
\lambda_a>1>\lambda_b.
\end{equation}
By virtue of $\eqref{lambda_ab}_1$, this leads to the restriction 
\begin{equation}
\rho\equiv\frac{R_1}{R_2}<\lambda_b<1\label{lambdaint}
\end{equation}
on $\lambda_b$, wherein we have defined, for later convenience, the notation $\rho$.
In this case, by \eqref{NM}$_1$, we must investigate the existence of positive roots for $\lambda_b$ in  the interval \eqref{lambdaint} of the equation
\begin{equation}
\int_{\lambda_b}^{\lambda_a}\frac{\hat{W}'(\lambda)}{\lambda^2}\mathrm{d}\lambda=0 \label{eq1}
\end{equation}
with $\lambda_a=\lambda_b/\rho$ for fixed $\rho$.

To this end, we introduce the function $f(\lambda_b)$ defined by 
\begin{equation}
 f(\lambda_b)=\int_{\lambda_b}^{\lambda_b/\rho}\frac{\hat{W}'(\lambda)}{\lambda^2}\mathrm{d}\lambda.\label{f}
\end{equation}
By virtue of \eqref{lambdaineqs}, we have
\begin{equation}
f(\rho)=\displaystyle\int_{\rho}^1\frac{\hat{W}'(\lambda)}{\lambda^2}\mathrm{d}\lambda<0, \quad
f(1)=\displaystyle\int_1^{1/\rho}\frac{\hat{W}'(\lambda)}{\lambda^2}\mathrm{d}\lambda>0,\label{boundaries}
\end{equation}
and by differentiation,
\begin{equation}
f'(\lambda_b)=\frac{1}{\rho}\frac{\hat{W}'\left(\lambda_a\right)}{\lambda_a^2} - \frac{\hat{W}'\left(\lambda_b\right)}{\lambda_b^2}.\label{first derivative}
\end{equation}
By \eqref{lambdaineqs} and \eqref{l1geql2}, this is clearly positive, and
we conclude that $f$ is  strictly increasing, and so $f$ has a unique zero, say $\lambda_b^*$, in the interval \eqref{lambdaint}. 
Consequently, a circular cylindrical sector made of an incompressible isotropic elastic material can  be straightened by applying terminal couples only. 
This result is universal to all constitutive models with strain-energy functions $\hat{W}$ continuously differentiable in $\mathbb{R}^+$ and satisfying the inequalities \eqref{lambdaineqs}.   The corresponding moment is
\begin{equation}
M^*=-\frac{HR_2^2{\lambda_b^*}^2}{2}\int_{\lambda_b^*}^{\lambda_b^*/\rho}\frac{\hat{W}'(\lambda)}{\lambda^4}\d \lambda.
\end{equation}
For illustration, we now report some analytical and numerical results for the solution of equation $\eqref{eq1}$. 

For the {\color{black}neo-Hookean} material \eqref{MR} we obtain
\begin{equation}
\lambda_b^*=\sqrt[4]{\frac{1-\rho^4}{4\ln (1/\rho)}}, \quad M^*=-\frac{\mu HR_2^2}{4}\left(1-\rho^2-\frac{1-\rho^6}{3{\lambda_b^*}^4}\right).
\end{equation}

For the Varga material, the explicit solution of $\eqref{eq1}$  and the corresponding moment are given by
\begin{equation}
\lambda_b^*=\sqrt{\displaystyle\frac{1+\rho+\rho^2}{3}}, \quad
 M^*=-\displaystyle\frac{\mu HR_2^2}{\lambda^*_b}\left(\frac{1-\rho^3}{3}-\frac{1-\rho^5}{5{\lambda_b^*}^2}\right).
\end{equation}

\begin{figure}[!h]
\centering	\includegraphics[width=\textwidth]{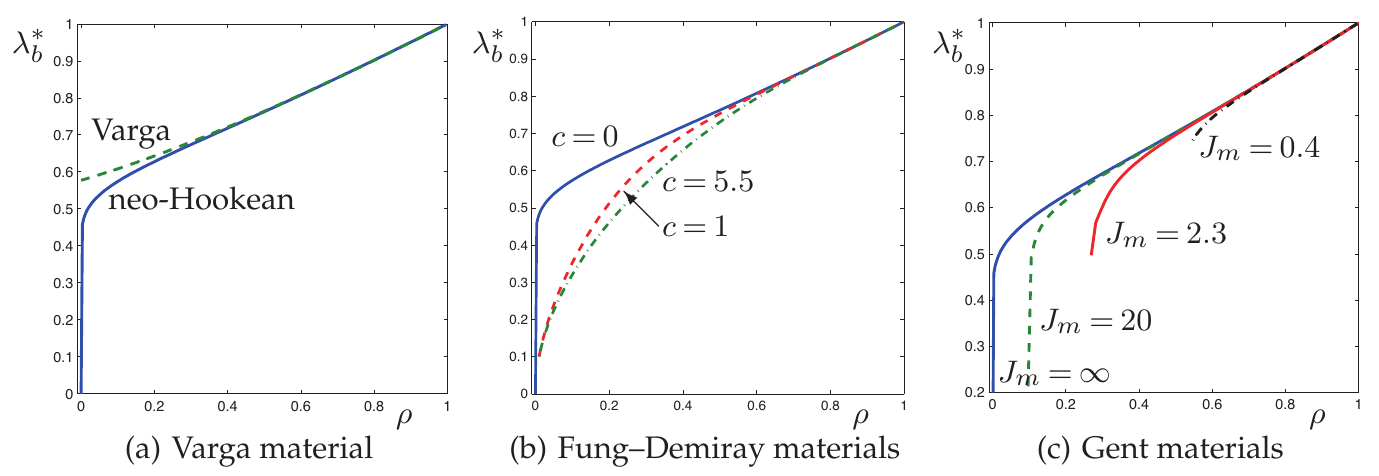}
		\caption{Circumferential stretch $\lambda_b^{*}$ as a function of the radii ratio $\rho=R_1/R_2$ for the straightening of blocks by end-couples only: (a) Varga and neo-Hookean materials;  (b)  Fung--Demiray  materials with stiffening parameters $c=5.5$, $c=1.0$, and $c \rightarrow 0$ (neo-Hookean limit);  (c)  Gent materials with stiffening parameters $J_m=0.4$, $J_m=2.3$, $J_m=20.0$, and $J_m \rightarrow \infty$ (neo-Hookean limit).\label{fig2}}
\end{figure}

For the Fung--Demiray material \eqref{F}, there are no explicit solutions, and a numerical resolution is required. Figures \ref{fig2}(a) and (b) display the stretch $\lambda_b^*$ as a function of the ratio $\rho=R_1/R_2$ for the neo-Hookean, Varga and Fung--Demiray materials. In particular, in plotting figure \ref{fig2}(b) we took the Fung--Demiray energy density with constants used in \cite{DeAC09} to model ``young'' human arteries ($c=1.0$) and ``old'' arteries ($c=5.5$).

As already pointed out, the assumptions that the function $\hat{W}$ is continuously differentiable and satisfies the inequalities \eqref{lambdaineqs} are fundamental  for proving the existence and uniqueness of the straightened configuration. We now show that, by means of slight changes, this result can be  extended to Gent materials \eqref{G}.
For these materials the function $\hat W$ reads
\begin{equation}
\hat{W}_{\mathrm{G}}(\lambda)=-\frac{\mu J_m}{2}\ln\left[1-\frac{(\lambda-\lambda^{-1})^2}{J_m}\right],
\end{equation}
and it  is continuously differentiable in the interval $(\lambda_m^{-1},\lambda_m)$  where  
\begin{equation} \label{lambda_m}
\lambda_m=\sqrt{\frac{J_m+2+\sqrt{J_m(J_m+4)}}{2}}
\end{equation}
is the upper bound on the stretch in (plane strain) uniaxial tension. Therefore, in order to straighten a circular cylindrical sector made of a Gent material, the circumferential stretch must belong to the interval $(\lambda_m^{-1},\lambda_m)$ throughout the thickness of the block. 
As a consequence of this restriction,  if 
$\lambda_m^{-2}<\rho< 1,
$ then the condition $\lambda_b\in(\lambda_m^{-1},\rho\lambda_m)$ implies that $\lambda\in(\lambda_m^{-1},\lambda_m)$ throughout the block. On the other hand, since $\hat{W}_{\mathrm{G}}$ satisfies the inequalities \eqref{lambdaineqs}, 
\begin{equation}
f_{\mathrm{G}}(\lambda_b):=\int_{\lambda_b}^{\lambda_b/\rho}\frac{\hat{W}'_{\mathrm{G}}(\lambda)}{\lambda^2}\mathrm{d}\lambda=\int_{\lambda_b}^{\lambda_b/\rho}\frac{\lambda^4-1}{\lambda^3(\lambda^2-\lambda_m^2)(\lambda^2-\lambda_m^{-2})}\mathrm{d}\lambda
\end{equation}
is an  increasing  function such that 
\begin{equation}
\lim_{\lambda_b\rightarrow\lambda_m^{-1}}f_{\mathrm{G}}(\lambda_b)=-\infty,\quad  f_{\mathrm{G}}(1)>0, \quad \lim_{\lambda_b\rightarrow\rho\lambda_m}f_{\mathrm{G}}(\lambda_b)=+\infty,
\end{equation}
and, if $\lambda_m^{-1}<\rho<1$, $f_{\mathrm{G}}(\rho)<0$.
 We may then conclude that $f_{\mathrm{G}}$  has a unique zero at 
 \begin{equation}\label{solutionGent}
 \lambda_b^*\in(\max\{\lambda_m^{-1},\rho\},\min\{1,\rho\lambda_m\}).
 \end{equation}
 It is worth noting that, in the light of \eqref{solutionGent},  $\lambda_b^*\rightarrow \lambda_m^{-1}$ as $\rho\rightarrow \lambda_m^{-2}$; see figure \ref{fig2}(c).

%%%%%%%%%%%%%%%%%%%%%%%%%%%%%%%

\subsection{Straightening by vice-clamps}

%%%%%%%%%%%%%%%%%%%%%%%%%%%%%%%

Now we investigate the existence of positive roots for $\lambda_b$ in the interval \eqref{lambdaint} when $M=0$, that is
\begin{equation}
\int_{\lambda_b}^{\lambda_a}\frac{\hat{W}'(\lambda)}{\lambda^4}\mathrm{d}\lambda=0.\label{eq2}
\end{equation}
Following  arguments similar to those used in the previous subsection, we introduce the function $g(\lambda_b)$ defined by 
\begin{equation}
 g(\lambda_b)=\int_{\lambda_b}^{\lambda_b/\rho}\frac{\hat{W}'(\lambda)}{\lambda^4}\d \lambda.\label{g}
\end{equation}
By virtue of \eqref{lambdaineqs}, we have
\begin{equation}
g(\rho)=\displaystyle\int_{\rho}^1\frac{\hat{W}'(\lambda)}{\lambda^4}\mathrm{d}\lambda<0, \quad
 g\left(1\right)=\displaystyle\int_1^{1/\rho}\frac{\hat{W}'(\lambda)}{\lambda^4}\mathrm{d}\lambda>0,\label{boundaries2}
\end{equation}
and by differentiation, 
\begin{equation}
g'(\lambda_b)=\frac{1}{\rho}\frac{\hat{W}'\left(\lambda_a\right)}{\lambda_a^4} - \frac{\hat{W}'\left(\lambda_b\right)}{\lambda_b^4}.\label{first derivative g}
\end{equation}
which, in view of \eqref{lambdaineqs} and \eqref{l1geql2},  is positive, implying that $g$ is strictly increasing. 
Therefore, by virtue of \eqref{boundaries2}, $g$ has a unique zero, say $\lambda_b^{**}$, in the interval $(\rho,1)$. Consequently, a circular cylindrical sector made of an incompressible isotropic elastic material can be straightened by applying a system of forces with zero resultant moment. By following the same arguments as in the previous section,  one can prove that this result is valid not only for  all the constitutive models with strain-energy functions $\hat{W}$ continuously differentiable in $\mathbb{R}^+$ and satisfying the inequalities \eqref{lambdaineqs},  but also for Gent materials. 
The corresponding resultant normal force is 
\begin{equation}
N^{**}=HR_2\lambda_b^{**}\int_{\lambda_b^{**}}^{\lambda_b^{**}/\rho}\frac{\hat{W}'(\lambda)}{\lambda^2}\d \lambda.
\end{equation}

\begin{figure}[!h]
	\centering
	\includegraphics[width=\textwidth]{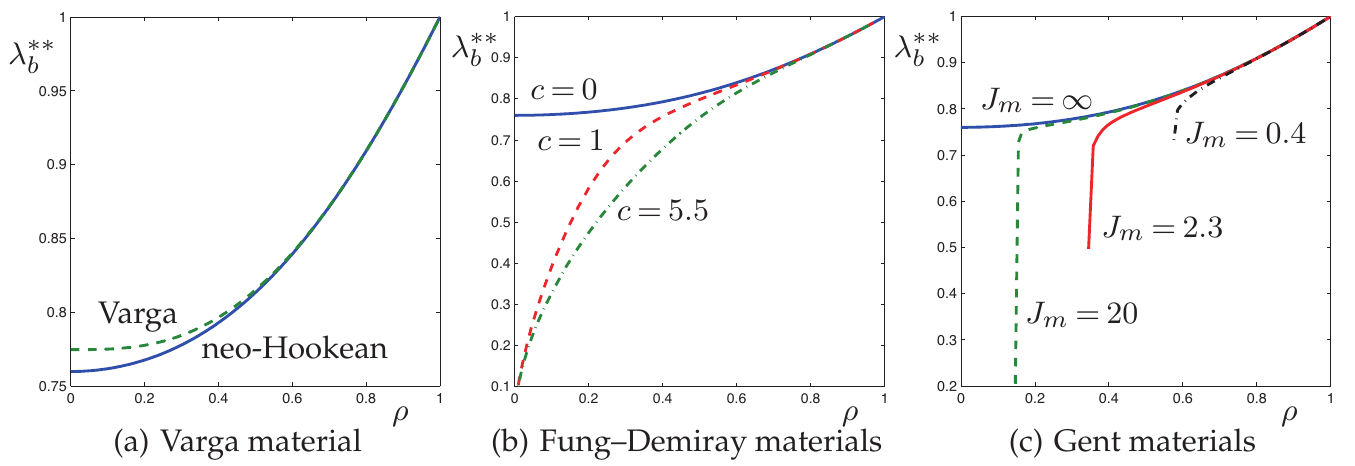}
		\caption{Circumferential stretch $\lambda_b^{**}$ as a function of the radii ratio $\rho=R_1/R_2$ for the straightening of blocks by vice-clamps: (a) Varga and neo-Hookean materials;  (b)  Fung--Demiray  materials with stiffening parameters $c=5.5$, $c=1.0$, and $c \rightarrow 0$ (neo-Hookean limit);  (c)  Gent materials with stiffening parameters $J_m=0.4$, $J_m=2.3$, $J_m=20.0$, and $J_m \rightarrow \infty$ (neo-Hookean limit). {\color{black} Note that different vertical scales are used in the three plots so as to avoid losing information. Thus, although not immediately apparent, the continuous curves in (a), (b) and (c) are the same and are for the neo-Hookean model.}\label{fig3}}
\end{figure}

For the neo-Hookean and Varga models the explicit solutions of \eqref{eq2}, which are illustrated in figure \ref{fig3}(a), and the corresponding total normal force are, respectively,
\begin{equation}
\lambda_b^{**}=\sqrt[4]{\frac{1+\rho^2+\rho^4}{3}},\quad
 N^{**}=\mu HR_2\lambda_b^{**}\displaystyle\left(\ln(1/\rho)-\frac{1-\rho^4}{{4\lambda_b^{**}}^4}\right),
\end{equation}
and
\begin{equation}
\lambda_b^{**}=\sqrt{\frac{3(1-\rho^5)}{5(1-\rho^3)}}, \quad
 N^{**}=2\mu H(R_2-R_1)\displaystyle\left(1-\frac{1+\rho+\rho^2}{3{\lambda_b^{**}}^2}\right).
\end{equation}
For the Fung--Demiray and Gent materials one can solve equation \eqref{eq2} only numerically. Figures \ref{fig3}(b) and (c) show $\lambda_b^{**}$ as a function of $\rho$ for different values of the material parameters $c$ and $J_m$.

We end this section by pointing out that, independently of the form of the strain-energy function, we have 
\begin{equation}\label{lambdasb}
\rho<\lambda_b^*<\lambda_b^{**}<1.
\end{equation}
We have already shown that $\lambda_b^*$ and $\lambda_b^{**}$ belong to the interval $(\rho,1)$.
Furthermore, from \eqref{lambdaineqs}, \eqref{f} and \eqref{g} we deduce that
\begin{equation}
g(\lambda_b)-f(\lambda_b)=-\int_{\lambda_b}^{\lambda_b/\rho}\frac{\lambda^2-1}{\lambda^4}\hat{W}'(\lambda)\mathrm{d} \lambda<0.
\end{equation}
Hence, since $f$ and $g$ are increasing in the interval $(\rho,1)$, the inequality \eqref{lambdasb} follows immediately.

%%%%%%%%%%%%%%%%%%%%%%%%%%%%%%%

\subsection{Thin sectors}

%%%%%%%%%%%%%%%%%%%%%%%%%%%%%%%

For thin sectors, that is sectors with thickness much smaller than the radius of the (undeformed) inner face, some general conclusions can be established about the straightened configuration.
For the asymptotic analysis we introduce the small parameter $\varepsilon>0$ defined as
\begin{equation}
\varepsilon= 1-\rho\ll1.\label{epsi}
\end{equation}

First we look at the straightening of a sector by the application of end couples, and we rewrite $f$ in \eqref{f} as a function of $\varepsilon$, specifically
\begin{equation}
F(\varepsilon)\equiv f(\lambda_b)=\int_{\lambda_b}^{\lambda_b/(1-\varepsilon)}\frac{\hat{W}'(\lambda)}{\lambda^2}\mathrm{d}\lambda.\label{new f}
\end{equation}
Expanding $F(\varepsilon)$ as a Maclaurin series in the parameter $\varepsilon$ up to the fifth order, substituting into the equation $f(\lambda_b)=0$ and dropping a common factor $\varepsilon$, yields the equation
\begin{eqnarray}
&&\hat{W}'\left(\lambda_b\right)+\frac{1}{2}\lambda_b\hat{W}''\left(\lambda_b\right)\varepsilon+\frac{1}{6}\left[2\lambda_b\hat{W}''\left(\lambda_b\right)+\lambda_b^2\hat{W}'''\left(\lambda_b\right)\right]\varepsilon^2\nonumber\\[0.5ex]
&&\quad+\frac{1}{24}\left[6\lambda_b\hat{W}''\left(\lambda_b\right)+6\lambda_b^2\hat{W}'''\left(\lambda_b\right)+\lambda_b^3\hat{W}^{\mathrm{iv}}\left(\lambda_b\right)\right]\varepsilon^3\nonumber\\[0.5ex]
&&\quad +\frac{1}{120}\left[24\lambda_b\hat{W}''\left(\lambda_b\right)+36\lambda_b^2\hat{W}'''\left(\lambda_b\right)+12\lambda_b^3\hat{W}^{\mathrm{iv}}\left(\lambda_b\right)+\lambda_b^4\hat{W}^\mathrm{v}\left(\lambda_b\right)\right]\varepsilon^4\nonumber\\[0.5ex]&&\quad+O(\varepsilon^5)=0.
\end{eqnarray}

Next, we expand $\lambda_b$ in terms of $\varepsilon$ to the fourth order:
\begin{equation}
\lambda_b=\lambda^{(0)}+\lambda^{(1)}\varepsilon+\lambda^{(2)}\varepsilon^2+\lambda^{(3)}\varepsilon^3+\lambda^{(4)}\varepsilon^4+O(\varepsilon^5).\label{lambdaaexpansion}
\end{equation}
Substituting this into the previous expansion and equating to zero the coefficients of each power in the resulting expression, we obtain, at zero order
\begin{equation}
\hat{W}'\left(\lambda^{(0)}\right)=0,
\end{equation}
and hence, by \eqref{lambdaineqs}, $\lambda^{(0)}=1$.
Using this result in the first-order term, we obtain
\begin{equation}
\left(\frac{1}{2}+\lambda^{(1)}\right)\hat{W}''\left(1\right)=0,
\end{equation}
and since $\hat{W}''\left(1\right)>0$, we deduce that $\lambda^{(1)}=-1/2$.  Then, the second-order term yields
\begin{equation}
\left(\lambda^{(2)}+\frac{1}{12}\right)\hat{W}''\left(1\right)+\frac{1}{24}\hat{W}'''\left(1\right)=0.
\end{equation}

The resulting expression for $\lambda_b$, to the second order in $\varepsilon$, is therefore
\begin{equation}
\lambda_b=1-\frac{1}{2}\varepsilon-\frac{1}{12}\left(1+\frac{1}{2}\frac{\hat{W}'''(1)}{\hat{W}''(1)}\right)\varepsilon^2+O(\varepsilon^3).\label{lambdaaWW}
\end{equation}
However, for plane strain, there is the universal result $\hat{W}'''(1)/\hat{W}''(1)=-3$ (see, for example, \cite{ogden1985}), so that the above formula reduces to
\begin{equation} 
\lambda_b=1-\frac{1}{2}\varepsilon+\frac{1}{24}\varepsilon^2+O(\varepsilon^3).
\end{equation}
Proceeding in a similar way (without showing the lengthy details), we obtain
\begin{equation} 
\lambda_b=1-\frac{1}{2}\varepsilon+\frac{1}{24}\varepsilon^2+\frac{1}{48}\varepsilon^3+\frac{1}{5760}\left[427-\frac{46\hat{W}^{\mathrm{iv}}(1)+3\hat{W}^{\mathrm{v}}(1)}{\hat{W}''(1)}\right]\varepsilon^4+O(\varepsilon^5).
\label{lambdab4order}
\end{equation}
Note, in particular, that the material properties do not enter until the fourth order, i.e. the results are independent of the form of strain-energy function up to order $\varepsilon^3$. 

Similarly, with an asymptotic analysis in the case of straightening by applying a resultant force only ($M=0$), we find that, for thin cylindrical sectors, the result analogous to \eqref{lambdab4order} is
\begin{equation} \label{exp2}
\lambda_b=1-\frac{1}{2}\varepsilon+\frac{5}{24}\varepsilon^2+\frac{5}{48}\varepsilon^3+\left[\frac{15}{128}-\frac{62\hat{W}^{\mathrm{iv}}(1) + 3\hat{W}^{\mathrm{v}}(1)}{5760\hat{W}''(1)}\right]\varepsilon^4+O(\varepsilon^5).
\end{equation}

The universal result used in \eqref{lambdab4order} and \eqref{exp2} can be confirmed by, for example, expanding the strain-energy function in terms of the Green strain tensor $\mathbf E = (\mathbf F^\text{T}\mathbf F - \mathbf I)/2$ in the form  
\begin{equation} \label{landau}
W=\mu \;  \tr(\mathbf{E}^2) + \dfrac{\mathcal A}{3}  \tr(\mathbf{E}^3) +\mathcal D \left(\tr(\mathbf{E}^2)\right)^2 + \ldots,
\end{equation}
where $\mu$, $\mathcal A$ and $\mathcal D$ are the second-, third-, and fourth-order elastic constants, respectively (see Destrade and Ogden \cite{DeOg10} and references therein). For the plane strain specialisation we have
\begin{equation}
\tr(\mathbf{E}^2) =\frac{1}{4}\left[(\lambda^2-1)^2+(\lambda^{-2}-1)^2\right],\quad \tr(\mathbf{E}^3) =\frac{1}{8}\left[(\lambda^2-1)^3+(\lambda^{-2}-1)^3\right],
\end{equation}
and by computing the successive derivatives of $\hat{W}$ we obtain
\begin{align}
& \hat{W}'(1) = 0, \quad \hat{W}''(1)= 4\mu, \quad \hat{W}'''(1) = -12\mu, \notag \\
& \hat{W}^\mathrm{iv}(1) = 156 \mu + 48 \mathcal{A} + 96\mathcal{D}, 
\quad \hat W^\mathrm{v}(1)=-120(11\mu+4\mathcal{A}+8\mathcal{D}).
\label{reduction}
\end{align}
Note that we established the last identity by expanding $W$ to one order further than in \eqref{landau}; see Ogden \cite{Ogde74}.
However, the next order terms do not contribute to the expression for $ \hat W^\mathrm{v}(1)$.

%%%%%%%%%%%%%%%%%%%%%%%%%%

\section{Incremental stability\label{sec-4}}

%%%%%%%%%%%%%%%%%%%%%%%%%%

We now study the stability of the deformed rectangular configuration by considering a superimposed incremental displacement $\mathbf{u}$, with components $(u_1,u_2,u_3)$.  
We denote the displacement gradient $\grad\mathbf{u}$ by $\mathbf{L}$, which has components $L_{ij}=\partial u_i/\partial x_j$.
When linearized the incremental incompressibility condition reads
\begin{equation}\label{incremental incompressibility}
\tr\mathbf{L}\equiv L_{ii}=u_{i,i}=0
\end{equation}
in the usual summation convention for indices, where a subscript $i$ following a comma signifies differentiation with respect to $x_i$.

The corresponding linearized incremental nominal stress referred to the deformed configuration, denoted $\mathbf{\dot{s}}_0$, is given by \cite{Ogde97}
\begin{equation}\label{incremental stress}
\mathbf{\dot{s}}_0=\boldsymbol{\mathcal{A}}_0\mathbf{L}+p\mathbf{L}-\dot{p}\mathbf{I},
\end{equation}
where a superposed dot signifies an increment, the zero subscript indicates evaluation in the deformed configuration, $\mathbf{I}$ is the identity tensor and $\boldsymbol{\mathcal{A}}_0$ is the fourth-order tensor of instantaneous elastic moduli.  In components, this reads
\begin{equation}
\dot{s}_{0ij}=\mathcal{A}_{0ijkl}u_{l,k}+pu_{i,j}-\dot{p}\delta_{ij},
\end{equation}
where $\delta_{ij}$ is the Kronecker delta.  Referred to the Eulerian principal axes of the underlying deformation, the only non-trivial components of $\boldsymbol{\mathcal{A}}_0$ are given by \cite{Ogde97}
\begin{equation}
\mathcal{A}_{0iijj}=\lambda_i\lambda_jW_{ij},\quad i,j\in\{1,2,3\},
\end{equation}
\begin{equation}
\mathcal{A}_{0ijij}=\mathcal{A}_{0ijji}+\lambda_iW_i=\frac{\lambda_iW_i-\lambda_jW_j}{\lambda_i^2-\lambda_j^2}\lambda_i^2,\quad i\neq j,\quad \lambda_i\neq\lambda_j\label{Acomps}
\end{equation}
with $W_i=\partial W/\partial\lambda_i$, $W_{ij}=\partial^2W/\partial\lambda_i\partial\lambda_j$, noting the major symmetry $\mathcal{A}_{0piqj}=\mathcal{A}_{0qjpi}$.  For $\lambda_i=\lambda_j$ the specializations of \eqref{Acomps} can be obtained by taking the limit $\lambda_j\rightarrow\lambda_i$ but are not needed here.

For the neo-Hookean material \eqref{neoH} these reduce to
\begin{equation}
\mathcal{A}_{0iiii}=\mu\lambda_i^2=\mathcal{A}_{0ijij},\quad \mathcal{A}_{0iijj}=\mathcal{A}_{0ijji}=0,\quad i\neq j.\label{neoHmoduli}
\end{equation}

In the absence of body forces the \emph{incremental equilibrium equation} has the general form 
\begin{equation}
\div  \mathbf{\dot{s}}_0=\mathbf{0},\label{eq}
\end{equation}
but here we consider  plane incremental deformations with $u_3=0$ and $u_1$ and $u_2$ independent of $x_3$, in which case the linearized incremental incompressibility condition becomes
\begin{equation}\label{inc}
u_{1,1}+u_{2,2}=0.
\end{equation}
Furthermore, since $p$ and the deformation, and hence  the components of $\boldsymbol{\mathcal{A}}_0$, depend only on $x_1$, the components of equation \eqref{eq} reduce to
\begin{equation}\label{scalar}
\dot{s}_{011,1}+\dot{s}_{021,2}=0, \quad \dot{s}_{012,1}+\dot{s}_{022,2}=0, \quad \dot{p}_{,3}=0.
\end{equation}
Therefore, $\dot{p}$ is independent of $x_3$. From \eqref{incremental stress}, the components of the incremental nominal stress $\mathbf{\dot{s}}_0$ appearing in \eqref{scalar} are 
\begin{eqnarray}
\dot{s}_{011}&=&(\mathcal{A}_{01111}+p)u_{1,1}+\mathcal{A}_{01122}u_{2,2}-\dot{p},\\
\dot{s}_{012}&=&\mathcal{A}_{01212}(u_{1,2}+u_{2,1}),\\
\dot{s}_{021}&=&\mathcal{A}_{02121}u_{1,2}+(\mathcal{A}_{02121}-\sigma_2)u_{2,1},\\
\dot{s}_{022}&=&\mathcal{A}_{02211}u_{1,1}+(\mathcal{A}_{02222}+p)u_{2,2}-\dot{p},
\label{components}
\end{eqnarray}
in the second of which we have used the fact that $\sigma_1=0$.

We seek solutions  of the form
\begin{equation}\label{assumption}
\{u_1,u_2,\dot{p}\}=\{U_1(x_1),U_2(x_1),P(x_1)\}\mathrm{e}^{\im nx_2},
\end{equation}
where $n=k\pi A/\Theta_0$ is the mode number and the integer $k$ is the \emph{number of wrinkles}.
Then,  the components of the incremental nominal stress \eqref{components}  have a similar form, which we write as
\begin{equation}
\dot{s}_{0ij}=S_{ij}(x_1)\mathrm{e}^{\im nx_2}, \quad i,j=1,2,
\end{equation}
with
\begin{eqnarray}
S_{11}&=&(\mathcal{A}_{01111}+p)U_1'+\im n\mathcal{A}_{01122}U_2-P,\\
S_{12}&=&\im n\mathcal{A}_{01212}U_1+\mathcal{A}_{01212}U_2',\label{S2}\\
S_{21}&=&\im n\mathcal{A}_{02121}U_1+(\mathcal{A}_{02121}-\sigma_2)U_2',\\
S_{22}&=&\mathcal{A}_{02211}U_1'+\im n(\mathcal{A}_{02222}+p)U_2-P,
\label{S}
\end{eqnarray}
and the  incremental incompressibility condition \eqref{inc} yields
\begin{equation}
U_1'=-\im nU_2.
\end{equation}
From \eqref{S2} we obtain
\begin{equation}
U'_2=-\im nU_1+\frac{S_{12}}{\alpha},
\end{equation}
where
\begin{equation}
\alpha=\mathcal{A}_{01212}=\frac{\lambda}{\lambda^4-1}\hat{W}'(\lambda).
\end{equation}
On use of the above equations followed by elimination of $S_{21}$ and $S_{22}$ in favour of $U_1$, $U_2$, $S_{11}$ and $S_{12}$, the incremental equilibrium equations \eqref{scalar} yield expressions for $S'_{11}$ and $S_{12}'$ in terms of $U_1$, $U_2$, $S_{11}$ and $S_{12}$. 

Then, by introducing the four-component displacement--traction vector  $\boldsymbol{\eta}=[U_1,U_2,\im S_{11},\im S_{12}]^\mathrm{T}$, we can cast the governing equations in the \emph{Stroh form}
\begin{equation}\label{stroh}
\frac{\mathrm{d}}{\mathrm{d} x_1}\boldsymbol{\eta}(x_1)=\im\mathbf{G}(x_1)\boldsymbol{\eta}(x_1),
\end{equation}
where the real Stroh matrix $\mathbf{G}$ has the form
\begin{equation}\label{gi}
\mathbf{G}=\left(\begin{array}{cccc}
0 & -n & 0 & 0\\
[3mm]
-n & 0 & 0 & -1/\alpha\\
[3mm]
n^2\sigma_2 & 0 & 0 & -n\\
[3mm]
0 & n^2 \nu & -n & 0
\end{array}\right),
\end{equation}
with
\begin{equation}
\nu = \mathcal{A}_{01111}+\mathcal{A}_{02222} + 2\mathcal{A}_{01212}-2\mathcal{A}_{01122}-2\mathcal{A}_{01221}=\lambda^2\hat{W}''(\lambda).
\end{equation}

We consider the incremental traction to vanish on the inner and outer faces, i.e.
\begin{equation}\label{bcxi}
S_{11}= S_{12}=0 \quad \mathrm{on}\quad x_1=a,b.
\end{equation}
If a solution of the incremental equations can be found subject to these boundary conditions then possible equilibrium states exist in a neighbourhood of the straightened configuration, signalling the onset of \emph{instability} of that configuration. The value of $1/(AR_2)=\lambda_b$ at this point is referred to as the \emph{critical value} for the stretch and denoted by $\lambda_\text{cr}$. Then, from \eqref{abR1R2} it follows that 
\begin{equation}
A(b-a)=\frac{1-\rho^2}{2\lambda_\text{cr}^2},
\end{equation}
which allows for the complete determination of the straightened geometry just prior to instability. In particular, the lengths of the block in the $x_1$ and $x_2$ directions are, respectively, given by
\begin{equation}\label{dimensions}
b-a=\frac{1-\rho^2}{2\lambda_\text{cr}}R_2,\quad 2l=2\Theta_0\lambda_\text{cr}R_2.
\end{equation}
Note that if $\lambda_b^*>\lambda_\text{cr}$, where $\lambda_b^*$ is the unique positive solution of \eqref{eq1}, then the cylindrical sector can be straightened by applying a moment alone without encountering any instability phenomenon. Similarly for a cylindrical sector straightened by vice-clamps, if  $\lambda_b^{**}>\lambda_{\mathrm{cr}}$ (this case is illustrated in figure \ref{fig-neoH} for a neo-Hookean material).

%%%%%%%%%%%%%

\section{Numerical results\label{sec-5}}

%%%%%%%%%%%%

In this section we investigate the possibility of solving numerically the incremental instability problem.
The Stroh form of the governing equations is numerically stiff and calls for the implementation of a robust algorithm.
In the incremental stability literature, the \emph{Compound Matrix Method} has been used successfully to solve a variety of stiff problems, including eversion  \cite{HaOr97,Fu02} and compression \cite{DoHa06} of cylindrical tubes, bending  \cite{CoDe08, DeAC09, RoBG11} and combined bending and compression \cite{Haug99} of a straight block, bending of a sector \cite{DeMO10}, and pressurisation of a spherical shell \cite{Fu98}. 
It turns out that for the problem considered here the compound matrix is itself singular, a  feature that seems to be unique to the straightening stability problem. 
We manage to circumvent this problem by constructing a reduced, non-singular, compound matrix.  
We then use the \emph{Impedance Matrix method}, which proves to be more precise numerically for this problem.
It also provides for a complete field description of the incremental displacement solution.

%In passing, we note that prior to the implementation of these shooting-like methods, we had tried to solve numerically the incremental stability problem as a boundary value problem. In order to use the \texttt{bvp4c} Matlab routine (based on the collocation method), we reformulated the original BVP, for which the location of the boundaries depended on $\lambda_\text{cr}$, into the ``standard'' form \cite{AsRu81} with fixed boundaries. 
%However, the resulting code failed to produce meaningful results and this line of computation had to be abandoned. 
%
%%%%%%%%%%%%%

\subsection{Compound matrix method}

%%%%%%%%%%%%%

Let $\boldsymbol{\eta}^{(1)}(x_1)$, $\boldsymbol{\eta}^{(2)}(x_1)$ be two linearly independent solutions of  \eqref{stroh}, and from them
generate the six compound functions $\phi_i,\,i\in\{1,\dots,6\}$, defined by
\begin{equation}
\phi_1=\left|\begin{array}{cc}
\eta_1^{(1)} & \eta_1^{(2)}\\
[2mm]
\eta_2^{(1)} & \eta_2^{(2)}
\end{array}\right|,\quad 
\phi_2=\left|\begin{array}{cc}
\eta_1^{(1)} & \eta_1^{(2)}\\
[2mm]
\eta_3^{(1)} & \eta_3^{(2)}
\end{array}\right|,\quad
\phi_3=\im\left|\begin{array}{cc}
\eta_1^{(1)} & \eta_1^{(2)}\\
[2mm]
\eta_4^{(1)} & \eta_4^{(2)}
\end{array}\right|,
\end{equation}
\begin{equation}
\phi_4=\im\left|\begin{array}{cc}
\eta_2^{(1)} & \eta_2^{(2)}\\
[2mm]
\eta_3^{(1)} & \eta_3^{(2)}
\end{array}\right|,\quad
\phi_5=\left|\begin{array}{cc}
\eta_2^{(1)} & \eta_2^{(2)}\\
[2mm]
\eta_4^{(1)} & \eta_4^{(2)}
\end{array}\right|,\quad
\phi_6=\left|\begin{array}{cc}
\eta_3^{(1)} & \eta_3^{(2)}\\
[2mm]
\eta_4^{(1)} & \eta_4^{(2)}
\end{array}\right|.
\end{equation}
Now, computing the derivatives of $\phi_i$ with respect to $x_1$ yields the so-called compound equations 
\begin{equation}\label{compound equations}
\frac{\d \boldsymbol{\phi}}{\d x_1}=\mathbf{A}(x_1)\boldsymbol{\phi}(x_1),
\end{equation}
where $ \boldsymbol{\phi}=(\phi_1,\dots,\phi_6)^\mathrm{T}$ and $\mathbf{A}$, the \emph{compound matrix}, has the form
\begin{equation}
\mathbf{A}=\left(\begin{array}{cccccc}
0 & 0 &-1/\alpha & 0 &0 &0 \\
0 & 0 & -n & -n & 0 & 0 \\
-n^2 \nu & n & 0 & 0 &n & 0\\
n^2\sigma_2 & n & 0 &0 &n & -1/\alpha\\
0 & 0& -n & -n & 0 & 0\\
0 & 0 & n^2\sigma_2 & - n^2\nu & 0 & 0
\end{array}\right).\label{compound matrix}
\end{equation}
The compound equations \eqref{compound equations} must be integrated numerically, starting with the initial condition
\begin{equation}
\boldsymbol{\phi}(a)=\phi_1(a)[1,0,0,0,0,0]^\mathrm{T},\label{init}
\end{equation}
and aiming at the target condition
\begin{equation}
\phi_6(b)=0,\label{target}
\end{equation}
according to \eqref{bcxi}.

However, we observe that $\det \mathbf{A}=0$, the first case in solid mechanics to our knowledge where a compound matrix is singular. 
Because of this situation, the numerical integration of the Cauchy problem \eqref{compound equations}--\eqref{init} will obviously encounter severe difficulties. 
In any event, the difficulty can be by-passed by noting that from \eqref{compound equations}--\eqref{init} it follows that
\begin{equation}\label{phi25}
\phi_2=\phi_5,\quad 
\frac{\d\phi_6}{\d x_1}=-n^2\alpha (\sigma_2+\nu)\frac{\d\phi_1}{\d x_1}+n\nu\frac{\d\phi_2}{\d x_1}.
\end{equation}
It then follows that we can construct a \emph{reduced compound matrix} $\mathbf{\hat{A}}$ by introducing five reduced compound functions $\psi_i,\,1\in\{1,\dots,5\}$, defined by
\begin{equation}\label{phi6}
\psi_i=\phi_i,\, (i=1,2,3,4), \quad \psi_5=\phi_6+n^2\alpha (\sigma_2+\nu)\phi_1 - n\nu \phi_2,
\end{equation}
so that from \eqref{compound equations} and \eqref{phi25}$_2$, the governing equations can be written as
\begin{equation}\label{non singular}
\frac{\d\boldsymbol{\psi}}{\d x_1}=\mathbf{\hat{A}}(x_1)\boldsymbol{\psi}(x_1),
\end{equation}
where $ \boldsymbol{\psi}=(\psi_1,\dots,\psi_5)^\mathrm{T}$ and $\mathbf{\hat{A}}$ has the form
\begin{equation}\label{b}
\mathbf{\hat{A}}=\left(\begin{array}{cccccc}
0 & 0 &-1/\alpha & 0 &0  \\
0 & 0 & -n & -n & 0  \\
-n^2\nu & 2n & 0 & 0  & 0\\
n^2(2\sigma_2+\nu) &  n(2-\nu/\alpha) & 0 &0  & -1/\alpha\\
f_1 & f_2 & 0  & 0 & 0
\end{array}\right).
\end{equation}
Here, 
\begin{equation}\label{ff}
f_1=n^2\frac{\d}{\d x_1}\left[\alpha (\sigma_2 + \nu)\right], \qquad f_2=- n\frac{\d \nu}{\d x_1}.
\end{equation}
It is easy to check that $\det\mathbf{\hat{A}}\neq0$. Hence, we may now integrate numerically the non-singular initial value problem \eqref{non singular}--\eqref{b} instead of the original compound equations with singular Jacobian. Finally,
in view of \eqref{init} and \eqref{phi6}, the initial condition for this system is
\begin{equation} \label{targ3}
\boldsymbol{\psi}(a)=\psi_1(a)\left(1,0,0,0,n^2\alpha(a) \left[\sigma_2(a)+ \nu(a)\right]\right)^\mathrm{T},
\end{equation}
and in view of \eqref{target} and $\eqref{phi6}_2$, the target condition is
\begin{equation}\label{new target}
\psi_5(b)=n^2\alpha(b) \left[\sigma_2(b)+ \nu(b)\right]\psi_1(b)- n\nu(b)\psi_2(b).
\end{equation}

To implement the (reduced) Compound Matrix method, we first non-dimensionalized equations \eqref{non singular}--\eqref{targ3} and specialised them to the neo-Hookean model. 
Then we applied an initial value solver (\texttt{ode45} or \texttt{ode15s} routines in Matlab), together with the dichotomy method  in order to satisfy the target condition \eqref{new target}. 

This approach is a shooting-like technique for which convergence and stability can depend on the well- or ill-conditioning of the underlying boundary value problem and of the target root finder problem.
The numerical drawbacks and challenges of the shooting techniques are highlighted in many textbooks; see for example \cite{asher}.

In our case, for a set of fixed values of $\rho=R_1/R_2 \in (0, 1)$, the bisection technique yields a sequence of values $\lambda_k$ to approximate $\lambda_{\mathrm{cr}}$ and stops the iterations when: ({a}) the residual $|F(\lambda_k)|=|\psi_5(b;\lambda_k)| \leq \mathrm{tol}_\mathrm{r}$ and ({b}) the error estimate $|\lambda_k-\lambda_{k-1}| \leq \mathrm{tol}_\mathrm{e}$. Both criteria must be used to check the goodness of the approximation. If the usual assumptions of the bisection method are satisfied on the starting localization interval $I_0 =[\lambda_0, \lambda_1]$ (that is $F(\lambda_0)F(\lambda_1) <0$) then the method will converge, i.e. by definition the criterion ({b}) will be always be satisfied. We set $\mathrm{tol}_\mathrm{e}=10^{-12}$ and $\mathrm{tol}_\mathrm{r} =10^{-4}$, and included a control on the maximum number of iterations allowed (it$_\text{max} = 40$).  We applied the same shooting approach when using the Compound Matrix method and the Impedance Matrix method (next section).
Both methods yielded the same results, although the former gave quite high residuals when $k>2$ and $\rho$ is small.  
However, the values of $\lambda_\text{cr}$ identified by the two methods were the same up to at least 4 significant digits even in the worst case of high residuals. 
For all intents and purposes, the Compound Matrix and the Impedance Matrix methods both provide the desired level of precision for the critical stretch of compression. 
The latter has the advantage of also providing a complete description of the incremental fields, as we now see.  

%%%%%%%%%%%%%

\subsection{Impedance matrix method}

%%%%%%%%%%%%%

Here we follow Shuvalov \cite{Shuv03} and introduce the matricant solution $\mathbf{M}(x_1,a)$ of \eqref{stroh}--\eqref{gi}
 defined as the matrix such that
\begin{equation}\label{matricant}
\boldsymbol{\eta}(x_1)=\mathbf{M}(x_1,a)\boldsymbol{\eta}(a), \quad \mathbf{M}(a,a)=\mathbf{I}_{(4)},
\end{equation}
which has the following $2 \times 2$ block structure
\begin{equation}\label{block}
\mathbf{M}(x_1,a)=\left(\begin{array}{cc}
\mathbf{M}_1(x_1,a) & \mathbf{M}_2(x_1,a)\\
\mathbf{M}_3(x_1,a) & \mathbf{M}_4(x_1,a)
\end{array}\right),
\end{equation} 
$\mathbf{I}_{(4)}$ being the fourth-order identity matrix.

We now use the incremental boundary condition $\mathbf{S}(a)=\mathbf{0}$ in \eqref{matricant} and \eqref{block} to establish that
\begin{equation}\label{derivation}
\mathbf S(x_1)=\mathbf{Z}_a(x_1)\mathbf{U}(x_1),\quad \textrm{where}\quad \mathbf{Z}_a=-\im\mathbf{M}_3\mathbf{M}_1^{-1}
\end{equation}
is the \emph{conditional impedance matrix}.
Substituting this impedance matrix into the incremental equilibrium equations  \eqref{stroh}--\eqref{gi} gives
\begin{equation}\label{intermediate}
\left\{\begin{array}{ll}
\displaystyle\frac{\mathrm{d} \mathbf{U}}{\mathrm{d}x_1}=\im\mathbf{G}_1\mathbf{U}-\mathbf{G}_2\mathbf{Z}_a\mathbf{U},\\
[3mm]
\displaystyle\frac{\mathrm{d} }{\mathrm{d}x_1}(\mathbf{Z}_a\mathbf{U})=\mathbf{G}_3\mathbf{U}+\im\mathbf{G}_1\mathbf{Z}_a\mathbf{U},
\end{array}\right.
\end{equation}
where
\begin{equation}
\mathbf{G}_1=\left(\begin{array}{cc}
0 & -n\\
-n & 0
\end{array}\right), \quad
\mathbf{G}_2=\left(\begin{array}{cc}
0 & 0\\
0 & -1/\alpha
\end{array}\right), \quad 
\mathbf{G}_3=\left(\begin{array}{cc}
n^2\sigma_2 & 0\\
0 & n^2\nu
\end{array}\right),
\end{equation}
are the $2\times 2$ sub-blocks of $\mathbf{G}$. Eliminating $\mathbf{U}$ between  the two equations in \eqref{intermediate} results in the following \emph{Riccati differential equation} for $\mathbf{Z}_a$:
\begin{equation}\label{riccati}
\frac{\mathrm{d}\mathbf{Z}_a}{\mathrm{d}x_1}=\im(\mathbf{G}_1\mathbf{Z}_a-\mathbf{Z}_a\mathbf{G}_1)+\mathbf{Z}_a\mathbf{G}_2\mathbf{Z}_a+\mathbf{G}_3.
\end{equation}
It is well behaved and can be integrated numerically in a robust way, subject to the initial condition,
\begin{equation} \mathbf{Z}_a(a)=\mathbf{0},
\end{equation}
which follows from \eqref{derivation}$_2$ and \eqref{matricant}$_2$.
The target condition is 
\begin{equation}
\det\mathbf{Z}_a(b)=0,
\end{equation}
which is met by adjustment of  the critical value $\lambda_\text{cr}$ for the stretch $\lambda_b$, by using a bisection approach as described in the previous section, with the same tolerance values for the stopping criteria.  
Then, $\mathbf S(b)=\mathbf{Z}_a(b)\mathbf{U}(b)=\mathbf{0}$ means that
\begin{equation}
\frac{U_2(b)}{U_1(b)}=-\frac{Z_{a11}(b)}{Z_{a12}(b)}=-\frac{Z_{a21}(b)}{Z_{a22}(b)},
\end{equation}
and this ratio determines the form of the wrinkles on the outer face of the straightened block.

To proceed, we introduce the \emph{dimensionless quantities}
\begin{equation}\label{dimless}
\left.\begin{array}{ll}
\displaystyle y=\frac{x_1}{b}\in[\rho^2,1], \quad n^\star=\frac{k\pi}{2\Theta_0}, \quad \alpha^\star=\frac{\alpha}{\mu},\quad \nu^\star=\frac{\nu}{\mu},\quad \sigma_2^\star=\frac{\sigma_2}{\mu}, \\
[5mm]
\displaystyle U_i^\star=\frac{U_i}{b}, \,i=1,2, \quad S_{1j}^\star=\frac{S_{1j}}{\mu},\,j=1,2, \quad \mathbf{Z}_a^\star=\frac{b}{\mu}\mathbf{Z}_a,
\end{array}\right.
\end{equation}
where $\mu= \frac{1}{4}\hat W''(1)$ is again the shear modulus in the reference configuration. 
Substitution of \eqref{dimless} into equations \eqref{stroh}--\eqref{gi} yields
\begin{equation}\label{dimless equation}
\frac{\mathrm{d}\boldsymbol{\eta}^\star}{\mathrm{d}y}=\im\mathbf{G}^\star\boldsymbol{\eta}^\star,
\end{equation}
with $\boldsymbol{\eta}^\star = [U_1^\star,U_2^\star,\im S_{11}^\star,\im S_{12}^\star]^\mathrm{T}$ and
\begin{equation}
\mathbf{G}^\star=\left(\begin{array}{cccc}
0 & -n^\star\lambda_\text{cr}^{-2} & 0 & 0\\
[3mm]
-n^\star\lambda_\text{cr}^{-2} & 0 & 0 & -1/\alpha^\star\\
[3mm]
n^{\star 2}\lambda_\text{cr}^{-4}\sigma_2^\star & 0 & 0 & -n^\star\lambda_\text{cr}^{-2}\\
[3mm]
0 & n^{\star 2}\lambda_\text{cr}^{-4}\nu^\star & -n^\star\lambda_\text{cr}^{-2} & 0
\end{array}\right).
\end{equation}

The dimensionless version of  the Riccati equation \eqref{riccati} is then
\begin{equation}\label{dimless riccati}
\frac{\mathrm{d}\mathbf{Z}_a^\star}{\mathrm{d}y}=\im(\mathbf{G}_1^\star\mathbf{Z}_a^\star-\mathbf{Z}_a^\star\mathbf{G}_1^\star)+\mathbf{Z}_a^\star\mathbf{G}_2^\star\mathbf{Z}_a^\star+\mathbf{G}_3^\star, \quad \mathbf{Z}_a^\star(\rho^2)=\mathbf{0},
\end{equation}
where
\begin{equation}
\mathbf{G}_1^\star=\left(\begin{array}{cc}
0 & -n^\star\lambda_\text{cr}^{-2}\\
-n^\star\lambda_\text{cr}^{-2} & 0
\end{array}\right), \quad
\mathbf{G}_2^\star=\left(\begin{array}{cc}
0 & 0\\
0 & -1/\alpha^\star
\end{array}\right)
\end{equation}
and 
\begin{equation}
\mathbf{G}_3^\star=\left(\begin{array}{cc}
n^{\star2}\lambda_\text{cr}^{-4}\sigma_2^\star & 0\\
0 & n^{\star 2}\lambda_\text{cr}^{-4}\nu^\star
\end{array}\right).
\end{equation}
The target condition to append to \eqref{dimless riccati} for finding the critical stretch $\lambda_\text{cr}$ is $\det\mathbf{Z}_a^\star(1)=0$. Finally, in order to determine the entire displacement field $\mathbf{U}$ throughout the straightened block  once equation \eqref{dimless riccati} is solved and the critical value $\lambda_\text{cr}$ has been found, we integrate  the equation
\begin{equation} \label{field}
\frac{\mathrm{d} \mathbf{U}^\star}{\mathrm{d}y}=\im\mathbf{G}_1^\star\mathbf{U}^\star - \mathbf{G}_2^\star \mathbf{Z}_a^\star \mathbf{U}^\star
\end{equation}
numerically (again using the \texttt{ode45} Matlab solver) from the initial conditions 
\begin{equation}
\frac{U_2^\star(1)}{U_1^\star(1)}=-\frac{Z_{a11}^\star(1)}{Z_{a12}^\star(1)}=-\frac{Z_{a21}^\star(1)}{Z_{a22}^\star(1)},
\end{equation}
at $y=1$ to the face at $y=\rho^2$.

%%%%%%%%%%%%%

\subsection{ Neo-Hookean materials}

%%%%%%%%%%%%%

For a neo-Hookean material with strain-energy function \eqref{neoH}, we obtain the non-dimensional quantities
\begin{equation}
\alpha^\star=\frac{y}{\lambda_\text{cr}^2},\quad \nu^\star = \frac{\lambda_\text{cr}^4+3y^2}{\lambda_\text{cr}^2y}, \quad \sigma_2^\star=\frac{\lambda_\text{cr}^4-y^2}{\lambda_\text{cr}^2y},
\end{equation}
which depend only on $y$ and $\lambda_\text{cr}$.

\begin{figure}
	\centering
	\includegraphics[width=\textwidth]{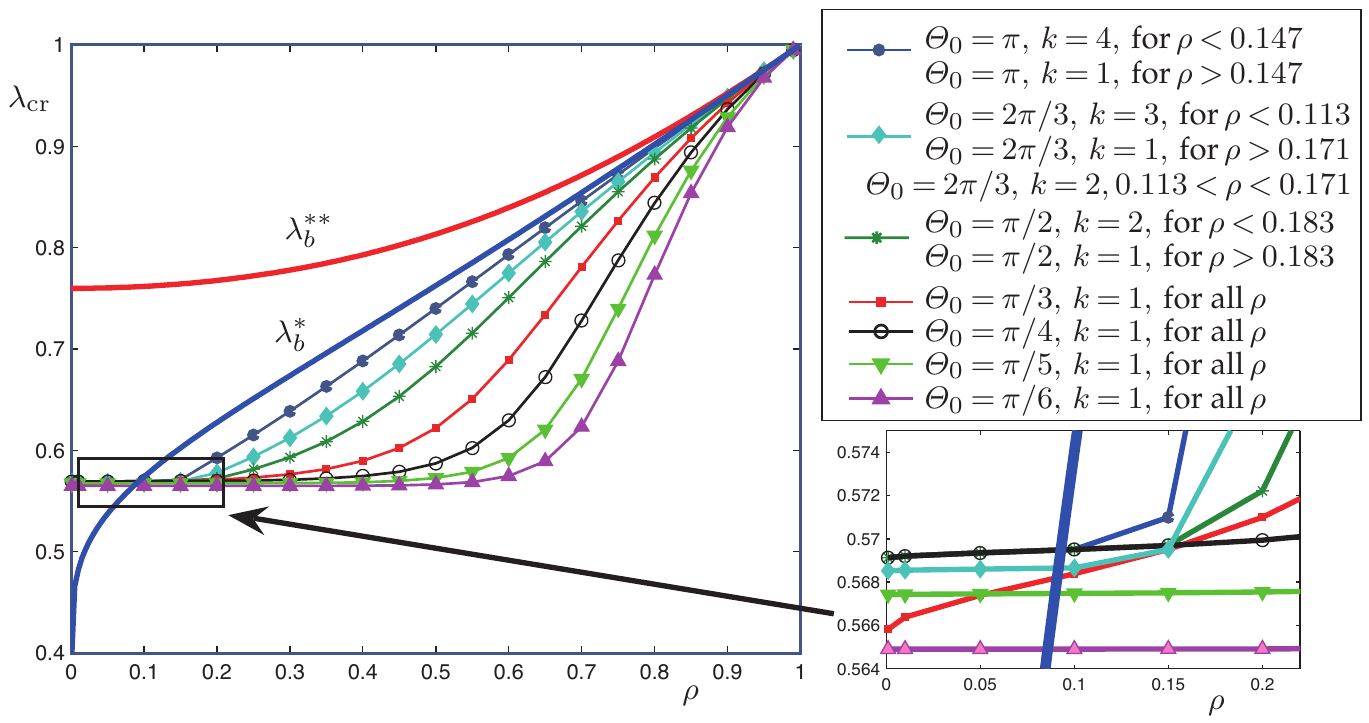}
	\caption{Critical value of the stretch $\lambda_{\mathrm{cr}}$ as a function of the radii ratio $\rho=R_1/R_2$ for the straightening of a neo-Hookean material, for different angles $\Theta_0$.}
	\label{fig-neoH}
\end{figure}

In figure \ref{fig-neoH} we provide plots of the critical value $\lambda_{\mathrm{cr}}$ versus $\rho=R_1/R_2$ corresponding to loss of stability of some straightened sectors of neo-Hookean materials. We take in turn $\Theta_0=\pi, 2\pi/3, \pi/2, \pi/3, \pi/4, \pi/5, \pi/6$. 
The number of wrinkles $k$ appearing on the compressed side of the block depends on the angle $\Theta_0$ and on  $\rho$. 
Hence, for $\Theta_0=\pi$ there are 4 wrinkles when $\rho < 0.1469$ and only one when $\rho > 0.1469$. 
For $\Theta_0=2\pi/3$ there are 3 wrinkles when $\rho < 0.1131$, 2 wrinkles when $0.1131 < \rho < 0.1717$, and one wrinkle when $\rho > 0.1717$. 
For $\Theta_0=\pi/2$ there are 2 wrinkles when $\rho < 0.1833$ and one wrinkle when $\rho > 0.1833$. 
For all the other values of the opening angle, there is only one wrinkle for any value of  $\rho$ in $(0,1)$. 
These results are summarised in the legend on the right of the figure.

In the thick sector--small wavelength limit  (i.e. as $\rho \rightarrow 0$ and $\Theta_0\rightarrow0$), we recover  the critical threshold for surface instability in plane strain of Biot \cite{Biot63} (i.e. $\lambda_\text{cr}\rightarrow 0.544$).

We also plot the curves for $\lambda_b^*$ and $\lambda_b^{**}$, giving the circumferential stretch of a cylindrical sector straightened by end couples and by vice-clamps, respectively. We see that a sector straightened by applying a system of forces and no moment (vice-clamps) never buckles on its outer face because $\lambda^{**}_b$ is always greater than $\lambda_\text{cr}$. 
However, a  sector straightened by moments alone and no normal forces (end couples) can buckle when $\rho$ is smaller than $0.09$ (see the zoom in figure \ref{fig-neoH}). 
When $\rho$  is greater than $0.09$, a  cylindrical sector straightened by end-couples does not present wrinkles on its outer face.

%%%%%%%%%%%%%

\subsection{ Gent materials}

%%%%%%%%%%%%%

\begin{figure}
	\centering
	\includegraphics[width=\textwidth]{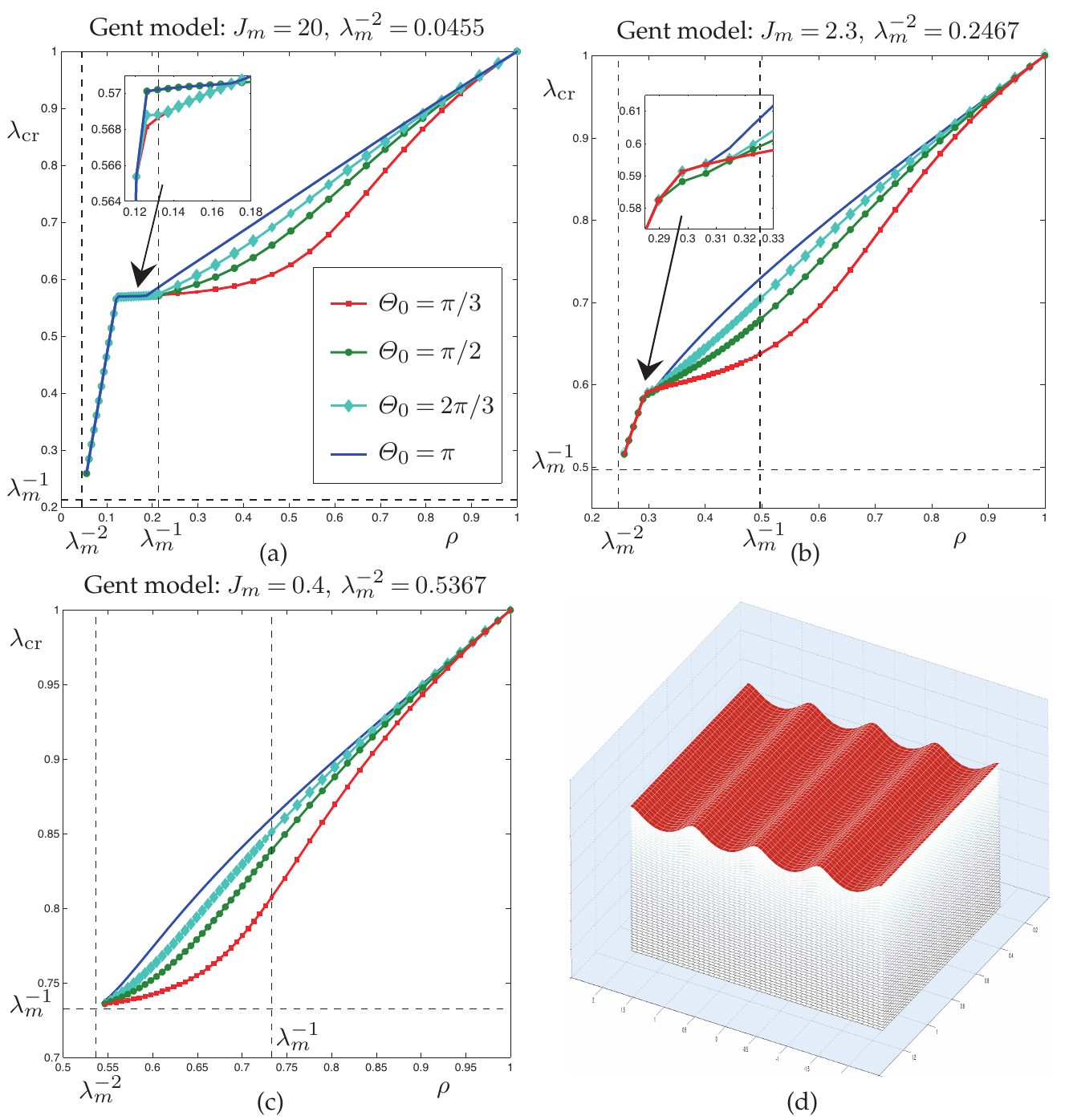}
		\caption{Instability of a straightened homogeneous circular cylindrical sector modelled by the  Gent strain-energy function. Figures (a)--(c) show the critical stretch $\lambda_{\mathrm{cr}}$ for wrinkling versus the radii ratio $\rho=R_1/R_2$ for different open angles $\Theta_0$, in the cases  where the Gent stiffening parameter is (a)  $J_m=20$ (rubber), (b) $J_m=2.3$ (old aorta) and (c) $J_m=0.4$ (young aorta). 
As summarised in table \ref{Table1}, the number of wrinkles $k$ depends on the constitutive parameter $J_m$, on the angle $\Theta_0$ and $\rho$.
For instance, for $J_m=20$, $\Theta_0 = \pi$ and $\rho=0.15$, the number of wrinkles is $k = 4$; see figures (a) and (d).\label{last-fig}}
\end{figure}

To investigate the influence of material parameters on the behaviour of straightened blocks, we use the Gent model \eqref{G}.
For its non-dimensional quantities in the Riccati equation \eqref{dimless riccati} we find
\begin{eqnarray}
\alpha^\star&=&\frac{J_my^2}{J_my\lambda_\text{cr}^2-(\lambda_\text{cr}^2-y)^2},\quad \nu^\star = \frac{2{\sigma_2^\star}^2}{J_m}+\frac{\sigma_2^\star(\lambda_\text{cr}^4+3y^2)}{\lambda_\text{cr}^4-y^2},\nonumber\\[12pt]
\sigma_2^\star&=&\frac{J_m(\lambda_\text{cr}^4-y^2)}{J_my\lambda_\text{cr}^2-(\lambda_\text{cr}^2-y)^2},
\end{eqnarray}
highlighting the role played by the stiffening parameter $J_m$, see the illustrations in figure \ref{last-fig}

Since a circular cylindrical sector made of a Gent material can be straightened provided that $\lambda_m^{-2}<\rho< 1$ and the circumferential stretch on the outer face $\lambda_b$  belongs to the interval $(\lambda_m^{-1},\rho\lambda_m)$, as indicated in \S \ref{sec-3}\ref{couples}, 
 $\lambda_\text{cr}$ tends to $\lambda_m^{-1}$ as $\rho\rightarrow\lambda_m^{-2}$.
Consequently, when $J_m$ is large enough but finite, the marginal stability curves for Gent and neo-Hookean materials are qualitatively similar in the interval $[\lambda_m^{-1},1)$, whereas they differ in the range $(\lambda_m^{-2},\lambda_m^{-1})$; compare, for instance, figure \ref{last-fig}(a) with figure \ref{fig-neoH}. 
As $J_m \rightarrow \infty$, the behaviour of neo-Hookean material is recovered. 

Finally, we integrated \eqref{field} for a case in which four wrinkles appear on the straightened face to generate the entire incremental displacement field (up to an arbitrary multiplicative factor), as illustrated in figure \ref{last-fig}(d).

\begin{table}[!t]
\begin{center}
    \begin{tabular}{|l|l|l|}
        \hline
  $J_m$ & Angle & Number of wrinkles \\
   \hline
   \hline
        20 & $\Theta_0=\pi/3$ & $k=1$ for $0.045 < \rho\le 1$ \\ \hline
        20 & $\Theta_0=\pi/2$ & $k=1$ for $0.045 < \rho < 0.11$ and  $0.21 < \rho \le 1$ \\
          ~ & ~ & $k=2$ for $0.11 < \rho < 0.21$  \\
          \hline
     20 & $\Theta_0=2\pi/3$ & $k=1$ for $0.045 < \rho < 0.11$ and  $0.21 < \rho \le 1$\\ 
        ~ & ~ & $k=2$ for $0.11 < \rho < 0.12$ and $0.13 < \rho < 0.21$ \\
          ~ & ~ & $k=3$ for $0.12 < \rho < 0.13$  \\
           \hline
      20& $\Theta_0=\pi$ & $k=1$ for $0.045 < \rho < 0.10$ and $0.19 < \rho\le 1$ \\ 
        ~ & ~ & $k=2$ for $0.10 < \rho < 0.11$ \\ 
        ~ & ~ & $k=3$ for $0.11 < \rho < 0.12$  and $0.17 < \rho < 0.19$\\ 
        ~ & ~ & $k=4$ for $0.12 < \rho < 0.17$ \\ 
        \hline
        \hline
        2.3 & $\Theta_0=\pi/3$ & $k=1$ for $0.25 < \rho \le 1$ \\ \hline
        2.3 & $\Theta_0=\pi/2$ & $k=1$ for $0.25 < \rho < 0.28$ and  $0.30 < \rho \le 1$ \\
           \hline
     2.3 & $\Theta_0=2\pi/3$ & $k=1$ for $0.25 < \rho < 0.28$ and  $0.31 < \rho \le 1$\\ 
        ~ & ~ & $k=2$ for $0.28 < \rho < 0.31$ \\
           \hline
      2.3& $\Theta_0=\pi$ & $k=1$ for $0.25 < \rho < 0.27$ and $0.31 < \rho \le 1$ \\ 
        ~ & ~ & $k=2$ for $0.27 < \rho < 0.28$ \\ 
        ~ & ~ & $k=3$ for $0.28 < \rho < 0.31$\\ 
      \hline
        \hline
        0.4 & $\Theta_0=\pi/3$ & $k=1$ for $0.54 < \rho \le 1$ \\ 
        \hline
        0.4 & $\Theta_0=\pi/2$ & $k=1$ for $0.54 < \rho \le 1$ \\
           \hline
     0.4 & $\Theta_0=2\pi/3$ & $k=1$ for $0.54 < \rho \le 1$\\ 
            \hline
      0.4& $\Theta_0=\pi$ & $k=1$ for $0.54 < \rho \le 1$ \\ 
        \hline
    \end{tabular}
    \caption{Description of the results displayed in figure \ref{last-fig} for the bifurcation curves of straightened Gent materials. There are 3 different types of Gent materials ($J_m=20$: rubber, $J_m=2.3$: young artery, $J_m=0.4$: old artery) and 4 different angles ($\Theta_0=\pi/3, \pi/2, 2\pi/3, \pi$).  Each curve is made of several pieces, each corresponding to the earliest bifurcation mode for a given value of $\rho$ of the sector, with corresponding number of wrinkles $k$ in the third column.}
    \label{Table1}
\end{center}
\end{table}

\section*{Acknowledgment}

Partial funding from the Royal Society of London (International Joint Project grant for MD, RWO, LV), from the Istituto Nazionale di Alta Matematica (Marie Curie COFUND Fellowship for LV;  GNCS Visiting Professor Scheme for MD, IS) and from the Ministero dell'Istruzione, dell'Universit\`a della Ricerca (PRIN-2009 project \emph{Matematica e
meccanica dei sistemi biologici e dei tessuti molli} for IS) is gratefully acknowledged.   

%%%%%%%%%%%%%%%%%

%%%%%%%%%% Insert bibliography here %%%%%%%%%%%%%%

\end{document}